\documentclass[aps,prl,twocolumn,10pt,showpacs,superscriptaddress]{revtex4}

\usepackage{graphicx}
\usepackage{amsmath,units}
\usepackage{mathrsfs}
\usepackage{color} 

\begin{document}
\title{Deep penetration fluorescence imaging through dense yeast cells suspensions using Airy beams}

\author{Harel Nagar}
	\affiliation{Raymond \& Beverly Sackler School of Chemistry, Tel Aviv
	University, Tel Aviv 6997801, Israel}
\author{Yael Roichman}
	\email{roichman@tauex.tau.ac.il}

\affiliation{Raymond \& Beverly Sackler School of Chemistry, Tel Aviv
	University, Tel Aviv 6997801, Israel}

\begin{abstract}
We propose a new method to image fluorescent objects through turbid media base on Airy beam scanning.
This is achieved by using the non-diffractive nature of Airy beams, namely their ability to maintain their shape while penetrating through a highly scattering medium.
We show, that our technique can image fluorescent objects immersed in turbid media with higher resolution and signal to noise than confocal imaging. As proof-of-principle, we demonstrate imaging of 1$\mu$m sized fluorescent beads through a dense suspension of yeast cells with an attenuation coefficient of 51cm$^{-1}$ at a depth of 90$\mu$m.
Finally, we demonstrate that our technique can also provide the depth of the imaged object without any additional sectioning.
\end{abstract}

\maketitle
Non-invasive imaging techniques are important tools in the biological and medical sciences, since they allow to study and characterize samples and tissue without physical contact. However, the heterogeneity of tissues causes strong scattering of light posing a challenge for optical imaging through tissue \cite{Dunsby2003}. To overcome this problem, a large variety of techniques base on various principles for imaging through turbid media have been developed in recent years\cite{Ntziachristos2010,Dunsby2003}. 
For example, spatial correlation methods \cite{Bertolotti2012,Katz2014,Yang2014} perform mathematical manipulation on the recorded speckle pattern received after the light passed through a scattering sample. These methods achieve diffraction limited resolution. Other approaches measure the transfer matrix of the scattering medium \cite{Choi2011,Popoff2010,Katz2012} and used it to image an object through the same medium. However, both types of methods, spatial correlation and transfer matrix measurement, require a pre-imaging measurement of the scattering properties of the media through which imaging should be performed. This means that imaging can be done either in static or in slow changing turbid media. While several approaches were suggested to allow for imaging in fast changing turbid media \cite{Liu2015,Wang2015}, for example by using the temporal memory effect of speckle illumination \cite{Edrei2016}, imaging through highly dynamic scattering media remains a challenge. 

Laser scanning techniques, such as confocal microscopy and two-photon microscopy \cite{Stephens2003,Zipfel2003}, are widely used for in-vivo imaging since they provide high resolution and improved axial sectioning with high imaging specificity. Two-photon microscopy enables deeper penetration into tissue, by working with near infra-red light that is scattered and absorbed less by biological tissue in comparison to visible light \cite{Helmchen2005}. Imaging deep into scattering biological samples is important since most organisms, tissues, and bodily fluids are not transparent enough to image a fluorescent object through them in either of these techniques.

In a previous paper, we presented the advantage of using the Airy beam scanning technique (ABST) for label free imaging due to its high penetration depth \cite{Nagar2018}. Airy beams \cite{Siviloglou2007_Obs} belong to the family of accelerating beams, which are beams that preserve their shape while propagating along curved trajectories \cite{Bandres2013}. A remarkable property of these beams is their ability to maintain their shape after encountering an obstacle \cite{Bandres2013,Fahrbach2010}. This property allows them to penetrate deeper than Gaussian beams into scattering media, while remaining focused. Here, we demonstrate that the same principle used previously to image through scattering media using reflection contrast can be applied to work with fluorescence contrast. Specifically, we acquire two dimensional (2D) projections of a florescent sample through dynamically changing turbid media in conditions where confocal imaging fails.

The ABST does not require sophisticated laser systems and can be perform with lower laser powers as compared to two-photon microscopy  (several $\mu$W instead of several mW) . It is also capable of imaging beyond the focal depth of the microscope objective. Another advantage of the ABST is its ability to measure the depth of a florescent signal by scanning with two reference beams and without performing multiple sectioning that is required in confocal and two-photon microscopy.

\begin{figure}[h]
	\centering
	\includegraphics[width=\linewidth]{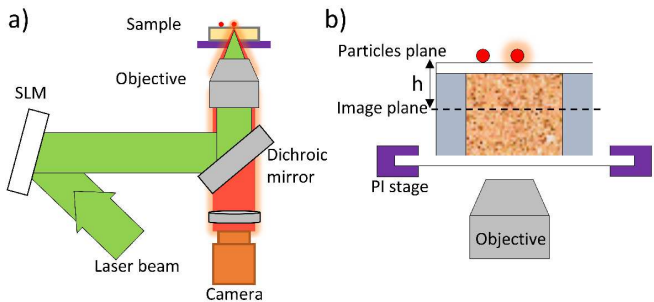}
	\caption{ (a) Schematic diagram of the imaging system. A laser beam covers the face of an SLM which encodes a cubic phase pattern on the beam. Then, the beam is inserted into an inverted microscope and is focused in a chamber filled with a yeast cell suspension. b) The laser beam focuses at a distance of $h$ from the fluorescence particles. Scanning is done by a piezoelectric microscope stage.}
	\label{fig:sys}
\end{figure}

The ABST setup is depicted in Fig.\ref{fig:sys}. A Gaussian laser beam  (Coherent, Verdi $\lambda=532$nm) is expanded and reflected from a spatial light modulator (SLM, Hamamatsu X10468-07). The SLM is used to transform the beam into a two dimensional Airy beam using a cubic phase mask \cite{Siviloglou2007_Obs}: $\phi=2\pi\ell(x^3+y^3)$, where $\phi$ is the phase, $x$ and $y$ are coordinates on the SLM normalized by the pixels size of the SLM, and $\ell$ is a scaling factor. 

The beam is then inserted into an inverted microscope (Olympus, IX71) and transformed by the objective into an Airy beam in the sample plane. In Fig.\ref{fig:sys} (b) the sample cell is depicted. It consists of a glass chamber $270\mu$m high filled with yeast cell suspensions (Saccharomyces Cerevisiae, Bravo instant yeast) in water with different attenuation coefficients $\mu$, as defined by the Beer-Lambert law $I/I_0=\exp(-\mu d)$  \cite{Mayerhofer2016}, with $d$ the penetration distance. On the far side of the sample cell one or two fluorescent beads (Silica $1.5\mu$m in diameter, Molecular Probes lots no. 1398541) are placed. The distance between the image plane and the particle plane $h$, is determined by translating the axial position of the objective in respect to the microscope stage. Scanning of the laser beam is achieved using a piezoelectric stage (PI-542.2CD) with high lateral resolution (up to 1 nm).

\begin{figure}[h]
	\centering
	\includegraphics[width=\linewidth]{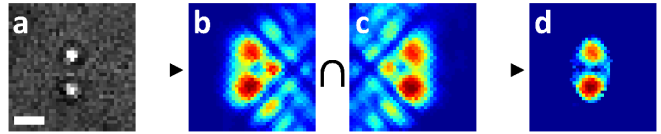}
	\caption{ Image reconstruction method. (a) Bright field image of two particles. (b,c) The same particles imaged by ABST with $\ell=30$ (b), and $\ell=-30$ (c).  (d) The ABST reconstructed image using the average of the intersection of the two images (b,c). Scale bar is $5\mu$m.}
	\label{fig:airy_analysis}
\end{figure}

Image reconstruction is done as follows, we illuminate a known spot on the far end of the sample using an Airy or Gaussian beam. If a florescent bead is present in that location the sample will be illuminated in the emission wavelength, otherwise it will remain dark. We measure the intensity of the fluorescent light in each location. We then construct a map of intensity at each location to create an image of the sample. The problem that arises for imaging fluorescent particles (Fig. \ref{fig:airy_analysis}a) out of the focal plane of the microscope is that the fluorescence signal is spread on a large area. This is different from the case of imaging reflected Airy beams, where the reflected beam stays focused\cite{Nagar2018}. Therefore, we adapt our scanning protocol to collect all the florescent signal. To this end, we use a relatively large window size for image reconstruction. Since the point spread function (PSF) of an Airy beam is not a single spot but includes side lobes, the obtained reconstructed image is a convolution of the shape of the sample and of the beam (see Fig. \ref{fig:airy_analysis}b). The PSF of a $+\ell$ Airy beam and a $-\ell$ Airy beam are mirror images of each other. Since they are curved in opposite directions, they reach the sample at a different location. The returning light from the reflecting or florescent sample is displace by an amount that depends on sample depth and a direction that depends on the sign of the topological charge of the beam. Imaging the same sample with two Airy beams $\ell,-\ell$ consecutively (Fig. \ref{fig:airy_analysis}b,c), allows us to decouple the signal from the main lobe from the rest of the image. To do so, we find the intersection area of both reconstructed images, and retain the average intensity of both images in this area only (Fig. \ref{fig:airy_analysis}d). Finally, we normalize the intensity of the image to obtain better contrast. In that manner we receive a new image with no artifacts from the side lobes. 

First, we compare the imaging capability of Gaussian and Airy beams at different $h$ through tap water. We use a beam intensity of $4\mu$W (at the sample plain) for imaging with a Gaussian beam and of  $60\mu$W for imaging with an Airy beam to unsure that the intensity of light in the main lobe of the Airy beam equals that of the Gaussian beam. We adjusted the window size at each $h$ to obtain maximum signal with as little noise as possible. This was done by choosing a window size slightly bigger than the returning fluorescent light diameter. The range of window sizes used was from $2\mu$m to $40\mu$m. Scanning was performed using a Gaussian beam, an Airy beam with $\ell=30$, and an Airy beam with $\ell=-30$ with a lateral step size of 0.5$\mu$m. We repeated the scans from $h=0$ to $h=90\mu$m with steps of $22.5\mu$m. 

\begin{figure}[h]
	\centering
	\includegraphics[width=\linewidth]{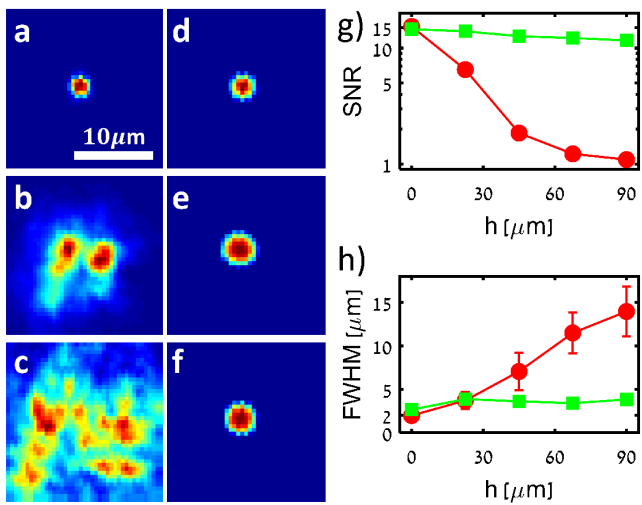}
	\caption{ a-f) Reconstructed images of one particle, with Gaussian beam based imaging (a-c) and with ABST (d-f) at different $h$: $h=0$  (a,d),  $h=45\mu$m (b,e) and $h=90\mu$m (c,f). SNR (g) and FWHM (h) as a function of $h$ for Gaussian (red circles) and Airy beam (green squares).}
	\label{fig:dif_h}
\end{figure}

The normalized reconstructed images for one particle at different depths are shown in Fig. \ref{fig:dif_h}a-f. Clearly the larger the distance of the particle from the imaging plane of the microscope is the less crisp the reconstructed images are. Using the ABST allows us to observe the particle even at a distance of $h=90\mu$m (Fig.~\ref{fig:dif_h}f), while imaging base on a Gaussian beam fails at depths as low as $h=45\mu$m (Fig.~\ref{fig:dif_h}b). To quantify these observations, we calculate the signal to noise ratio (SNR) and the full width at half maximum (FWHM) of the single particle reconstructed images (Fig.~\ref{fig:dif_h}g,h). The SNR was estimated by the ratio of the average intensity at the particle center and the average intensity far from the particle. The FWHM was taken from the intensity cross section along the image of the particle. The SNR of the ABST reduced slightly with $h$ at a rate of 0.0025$\mu$m$^{-1}$, while for the Gaussian beam based imaging it deteriorated rapidly at a rate faster than 0.03$\mu$m$^{-1}$. Close to the microscope focus, the image resolution as measured by the FWHM is higher for Gaussian beam based imaging. However, at a depth of $h=22.5\mu$m the FWHM of both techniques are the same and at larger depths ABST maintains the same FWHM, while for the  Gaussian beam based imaging the FWHM increases almost exponentially with a rate of approximately $0.026\mu$m$^{-1}$.

\begin{figure}[h]
	\centering
	\includegraphics[width=\linewidth]{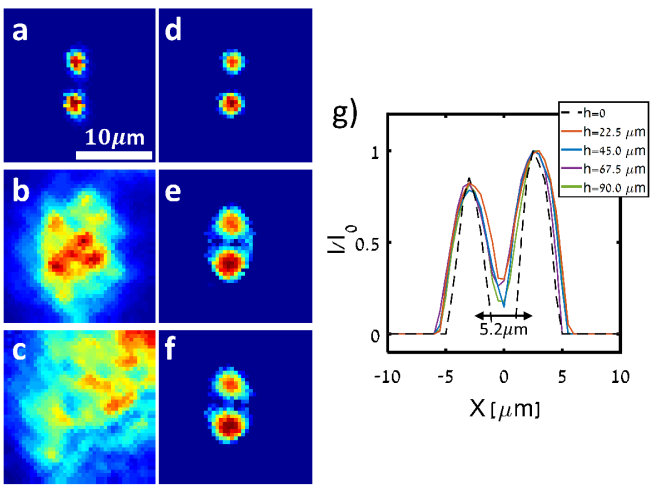}
	\caption{ a-f) Reconstructed images of two particles, with Gaussian beam based imaging (a-c) and with ABST (d-f) at different $h$: $h$=0  (a,d),  $h=45\mu$m (b,e) and $h=90\mu$m (c,f). g) Normalized intensity cross-section along the two particles as a function of $h$.}
	\label{new}
\end{figure}

Next, we estimate the imaging resolution of both techniques (Fig. \ref{new}a-g) by imaging two particles $5.2\mu$m apart at different $h$. We observed the same trends in the SNR and FWHM as in the one particle measurements, where the ABST outperformed the Gaussian beam imaging at large depths (Fig. \ref{new}a-f). The normalized intensity cross sections through the particles centers for ABST at different $h$ are shown in Fig. \ref{fig:dif_h}g. The separation in the cross section profile of the two particles is clearly observed for ABST in all depths $h$. For the Gaussian beam based imaging the particles are indistinguishable for $h$ larger than $45\mu$m. 

\begin{figure}[htbp]
 	\centering
 	\includegraphics[width=0.9\linewidth]{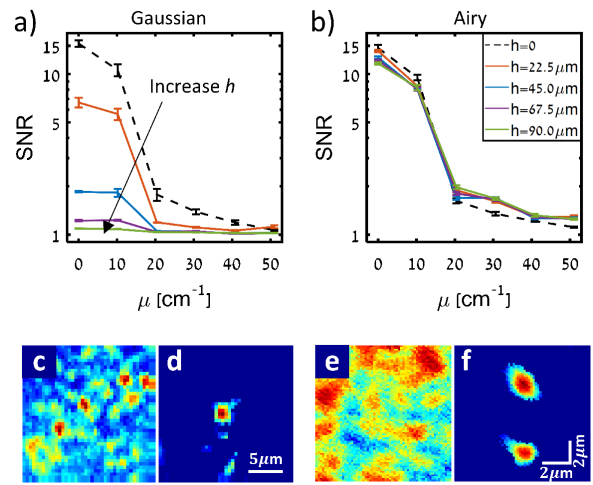}
 	\caption{Imaging through suspensions of yeast cells with different concentrations. (a,b) The SNR at different $h$ and yeast cell concentration for the Gaussian and Airy beam based imaging, respectively.  Reconstructed images of one particle at $h=90\mu$m and $\mu=51\text{cm}^{-1}$ for Gaussian beam based imaging (c) and ABST (d). (e,f) High resolution scanning, with 150nm lateral step size, of two particles through a yeast cells suspension at  $h=90\mu$m and $\mu=31\text{cm}^{-1}$ with Gaussian and Airy beam based imaging, respectively.}
 	\label{fig:turbid}
 \end{figure}

We now apply the ABST to imaging through turbid media. We repeat the previous experiments, imaging through suspensions of yeast cells with different concentration with an attenuation coefficient of up to $\mu=51\text{cm}^{-1}$. Yeast cells are eukaryote organisms that are used extensively in biological studies due to their similarity to human cells \cite{Beauvoit1993}. We find that for $h=0$ both techniques have the same SNR decay for increasing concentration. However, for larger $h$, the SNR of the Gaussian beam base imaging reduces dramatically with concentration (Fig. \ref{fig:turbid}a), while the ABST SNR remains almost the same (Fig. \ref{fig:turbid}b) due to its non-diffracting nature. The most dramatic effect can be seen for a one particle reconstructed image at $h=90\mu$m and $\mu=51\text{cm}^{-1}$ (Fig. \ref{fig:turbid}c,d), where ABST manages to image a single particle and the  Gaussian beam based imaging fails miserably. We also performed imaging at higher resolution using a smaller lateral step size of 150nm instead of 500nm (Fig. \ref{fig:turbid}e,f). Here we can see a much better separation between the particles. The FWHM of each particle is $1.7\mu$m compared to their diameter of $1.5\mu$m.

\begin{figure}[htbp]
	\centering
	\includegraphics[width=\linewidth]{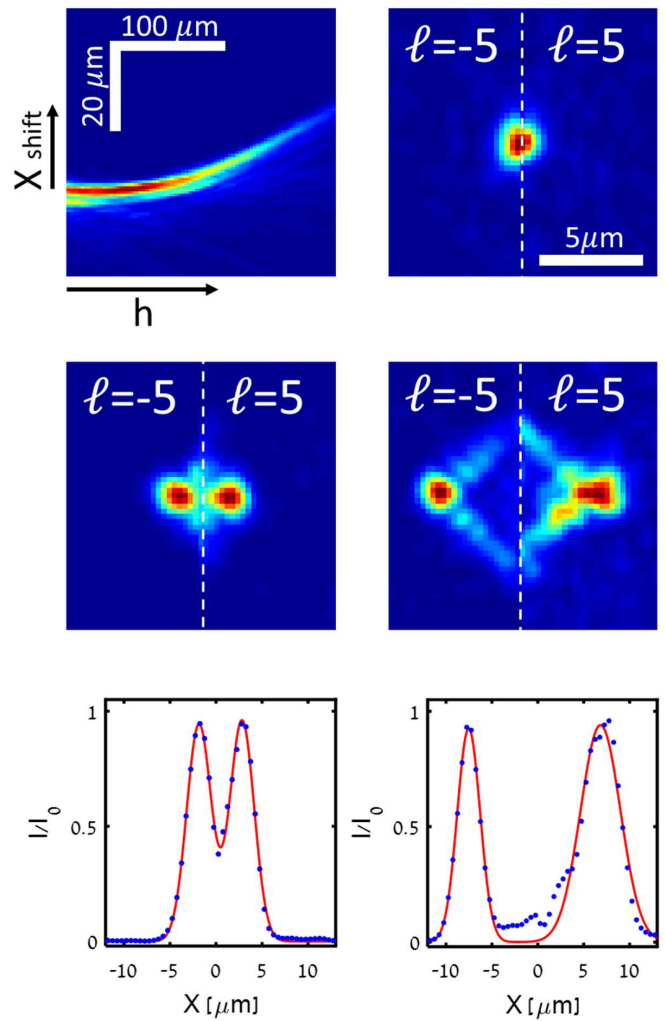}
	\caption{Depth measurements. (a) Volumetric imaging of an Airy beam of $\ell=5$. (b-d) Reconstructed image of fluorescent particles at depths of $h=0,75,135\mu$m, respectively, where the left side of the image is the reconstructed image made using an Airy beams of $\ell=-5$ and the the right side made using an Airy beam of $\ell=5$. (e,f) Normalized intensity cross-section of the main lobes of (c,d), respectively.}
	\label{fig:h_meas}
\end{figure}

Finally, we utilized the Airy beam curved trajectory for measuring the distance of fluorescence object from the focal plan. First, we characterize the shape of the main lobe of the Airy beam. We use volumetric imaging \cite{Roichman2006} to image  Airy beams of $\ell=5$ and $\ell=-5$ through water using a mirror (see Fig.~\ref{fig:h_meas}a). These beams have a relatively high curvature and decay significantly within the $270\mu$m depth of the sample. We then fit the measured main lobe trajectory in the axial direction to a parabolic function. The two beams ($\ell=5,-5$) bend to opposites directions but with almost the same absolute parabolic squared term coefficient. There is approximately 6\% deviation between the different beams. The relation between the displacement of the main lobe ($X_{\text{shift}}$) and $h$ for each beam was found to be: $X_{\text{shift}}=5\cdot 10^{-4} h^2$. 
We then prepared a sample with fluorescent beads at three different depths, $h=0,75,135\mu$m, and filled it with a 20\% milk/water mixture ($\mu=13\text{cm}^{-1}$). We then imaged the beads at the different depths with two reference Airy beams of $\ell=5$ and $\ell=-5$, acquiring two images. We know the center of the optical axis by projecting an Airy beam with $\ell=50$ that hardly curves at these length scales. We superpose the two images of the sample taken with the two beams relative to the optical axis and observe their relative shift. In Fig. \ref{fig:h_meas}b-d the right side of the image is the right part of the $\ell=5$ reconstructed image and the left side is the left part of the $\ell=-5$ image. Clearly, for $h=0$ the beams arrive at the same point on the sample and are imaged there, resulting in overlapping images (Fig. \ref{fig:h_meas}b). However, once $h$ is increased the main lobes appear separated, as they have shifted in opposite directions (Fig. \ref{fig:h_meas}c,d). We measured the normalized intensity along the center of the main lobes for $h=75,135\mu$m (Fig. \ref{fig:h_meas}c,d) and fitted it to two Gaussian functions shown in Fig. \ref{fig:h_meas}e,f, respectively. The distance between the two Gaussian peaks was measured and used to calculate $h$ using the parabolic trajectory fit obtained above. The calculated distances for $h=75\mu$m and $135\mu$m where $h_{calc}=70\pm5\mu$m and $123\pm7\mu$m, respectively. Agreeing with the known values.

In conclusion, we have demonstrated that we can use the ABST deep penetration advantages not only for label free imaging as we have demonstrated previously, but also for fluorescence imaging. We have implemented the ABST to image through dense suspensions of yeast cells and showed that it is superior to confocal-like imaging. We also showed that the curved shape of an Airy beam can be used for measuring the depth of the fluorescent object simply by scanning with two reference beams with no need for sectioning. 
The main advantage of using ABST over two photon microscopy is the low laser powers that it requires, which is important in reducing radiation damage to the investigated sample. The ability to get 3D data from only two scans is an added benefit, which will be mostly suitable for sparse samples. The technique is also compatible with live cell imaging since it is not affected by changes in the scattering media at all. The resolution of the technique depends mainly on the width of the main lobe of the Airy beam, and is therefore slightly lower than the diffraction limit. However, using super Airy beams \cite{Brijesh2015} for imaging might improve the resolution. Finally, using an array of Airy beams for parallel imaging will enable fast 3D imaging of live organisms.    
\newline
\newline
\noindent{ \bf Funding}. Chief scientist of Israel, Kamin project no. 55305 and Israel Science Foundation Grant no. 988/17.  
\newline
\newline

% Bibliography
%\bibliography{refs_paper}

% Full bibliography added automatically for Optics Letters submissions; the following line will simply be ignored if submitting to other journals.
% Note that this extra page will not count against page length

\end{document}